\RequirePackage{lineno}
\documentclass[aps,prl,twocolumn,superscriptaddress,showpacs,preprintnumbers,amsmath,amssymb]{revtex4}
\usepackage{graphicx} 
\usepackage{dcolumn}  

\graphicspath{{ps}}

\begin{document}


\title{ \quad\\[1.0cm] Observation of $\Xi(1620)^0$ and evidence for $\Xi(1690)^0$ in $\Xi_c^+ \rightarrow \Xi^-\pi^+\pi^+$ decays}

\noaffiliation
\affiliation{University of the Basque Country UPV/EHU, 48080 Bilbao}
\affiliation{Beihang University, Beijing 100191}
\affiliation{Brookhaven National Laboratory, Upton, New York 11973}
\affiliation{Budker Institute of Nuclear Physics SB RAS, Novosibirsk 630090}
\affiliation{Faculty of Mathematics and Physics, Charles University, 121 16 Prague}
\affiliation{Chonnam National University, Kwangju 660-701}
\affiliation{University of Cincinnati, Cincinnati, Ohio 45221}
\affiliation{Deutsches Elektronen--Synchrotron, 22607 Hamburg}
\affiliation{Duke University, Durham, North Carolina 27708}
\affiliation{University of Florida, Gainesville, Florida 32611}
\affiliation{Key Laboratory of Nuclear Physics and Ion-beam Application (MOE) and Institute of Modern Physics, Fudan University, Shanghai 200443}
\affiliation{Justus-Liebig-Universit\"at Gie\ss{}en, 35392 Gie\ss{}en}
\affiliation{Gifu University, Gifu 501-1193}
\affiliation{II. Physikalisches Institut, Georg-August-Universit\"at G\"ottingen, 37073 G\"ottingen}
\affiliation{SOKENDAI (The Graduate University for Advanced Studies), Hayama 240-0193}
\affiliation{Gyeongsang National University, Chinju 660-701}
\affiliation{Hanyang University, Seoul 133-791}
\affiliation{University of Hawaii, Honolulu, Hawaii 96822}
\affiliation{High Energy Accelerator Research Organization (KEK), Tsukuba 305-0801}
\affiliation{J-PARC Branch, KEK Theory Center, High Energy Accelerator Research Organization (KEK), Tsukuba 305-0801}
\affiliation{IKERBASQUE, Basque Foundation for Science, 48013 Bilbao}
\affiliation{Indian Institute of Science Education and Research Mohali, SAS Nagar, 140306}
\affiliation{Indian Institute of Technology Bhubaneswar, Satya Nagar 751007}
\affiliation{Indian Institute of Technology Guwahati, Assam 781039}
\affiliation{Indian Institute of Technology Hyderabad, Telangana 502285}
\affiliation{Indian Institute of Technology Madras, Chennai 600036}
\affiliation{Indiana University, Bloomington, Indiana 47408}
\affiliation{Institute of High Energy Physics, Chinese Academy of Sciences, Beijing 100049}
\affiliation{Institute of High Energy Physics, Vienna 1050}
\affiliation{Institute for High Energy Physics, Protvino 142281}
\affiliation{INFN - Sezione di Napoli, 80126 Napoli}
\affiliation{INFN - Sezione di Torino, 10125 Torino}
\affiliation{Advanced Science Research Center, Japan Atomic Energy Agency, Naka 319-1195}
\affiliation{J. Stefan Institute, 1000 Ljubljana}
\affiliation{Institut f\"ur Experimentelle Teilchenphysik, Karlsruher Institut f\"ur Technologie, 76131 Karlsruhe}
\affiliation{King Abdulaziz City for Science and Technology, Riyadh 11442}
\affiliation{Department of Physics, Faculty of Science, King Abdulaziz University, Jeddah 21589}
\affiliation{Korea Institute of Science and Technology Information, Daejeon 305-806}
\affiliation{Korea University, Seoul 136-713}
\affiliation{Kyoto University, Kyoto 606-8502}
\affiliation{Kyungpook National University, Daegu 702-701}
\affiliation{LAL, Univ. Paris-Sud, CNRS/IN2P3, Universit\'{e} Paris-Saclay, Orsay}
\affiliation{\'Ecole Polytechnique F\'ed\'erale de Lausanne (EPFL), Lausanne 1015}
\affiliation{P.N. Lebedev Physical Institute of the Russian Academy of Sciences, Moscow 119991}
\affiliation{Faculty of Mathematics and Physics, University of Ljubljana, 1000 Ljubljana}
\affiliation{Ludwig Maximilians University, 80539 Munich}
\affiliation{Luther College, Decorah, Iowa 52101}
\affiliation{University of Maribor, 2000 Maribor}
\affiliation{Max-Planck-Institut f\"ur Physik, 80805 M\"unchen}
\affiliation{School of Physics, University of Melbourne, Victoria 3010}
\affiliation{University of Mississippi, University, Mississippi 38677}
\affiliation{University of Miyazaki, Miyazaki 889-2192}
\affiliation{Moscow Physical Engineering Institute, Moscow 115409}
\affiliation{Moscow Institute of Physics and Technology, Moscow Region 141700}
\affiliation{Graduate School of Science, Nagoya University, Nagoya 464-8602}
\affiliation{Kobayashi-Maskawa Institute, Nagoya University, Nagoya 464-8602}
\affiliation{Universit\`{a} di Napoli Federico II, 80055 Napoli}
\affiliation{Nara Women's University, Nara 630-8506}
\affiliation{National Central University, Chung-li 32054}
\affiliation{National United University, Miao Li 36003}
\affiliation{Department of Physics, National Taiwan University, Taipei 10617}
\affiliation{H. Niewodniczanski Institute of Nuclear Physics, Krakow 31-342}
\affiliation{Nippon Dental University, Niigata 951-8580}
\affiliation{Niigata University, Niigata 950-2181}
\affiliation{Novosibirsk State University, Novosibirsk 630090}
\affiliation{Osaka City University, Osaka 558-8585}
\affiliation{Pacific Northwest National Laboratory, Richland, Washington 99352}
\affiliation{Panjab University, Chandigarh 160014}
\affiliation{Peking University, Beijing 100871}
\affiliation{University of Pittsburgh, Pittsburgh, Pennsylvania 15260}
\affiliation{Punjab Agricultural University, Ludhiana 141004}
\affiliation{Research Center for Nuclear Physics, Osaka University, Osaka 567-0047}
\affiliation{Theoretical Research Division, Nishina Center, RIKEN, Saitama 351-0198}
\affiliation{University of Science and Technology of China, Hefei 230026}
\affiliation{Showa Pharmaceutical University, Tokyo 194-8543}
\affiliation{Soongsil University, Seoul 156-743}
\affiliation{University of South Carolina, Columbia, South Carolina 29208}
\affiliation{Stefan Meyer Institute for Subatomic Physics, Vienna 1090}
\affiliation{Sungkyunkwan University, Suwon 440-746}
\affiliation{School of Physics, University of Sydney, New South Wales 2006}
\affiliation{Department of Physics, Faculty of Science, University of Tabuk, Tabuk 71451}
\affiliation{Tata Institute of Fundamental Research, Mumbai 400005}
\affiliation{Excellence Cluster Universe, Technische Universit\"at M\"unchen, 85748 Garching}
\affiliation{Department of Physics, Technische Universit\"at M\"unchen, 85748 Garching}
\affiliation{Department of Physics, Tohoku University, Sendai 980-8578}
\affiliation{Earthquake Research Institute, University of Tokyo, Tokyo 113-0032}
\affiliation{Department of Physics, University of Tokyo, Tokyo 113-0033}
\affiliation{Tokyo Institute of Technology, Tokyo 152-8550}
\affiliation{Tokyo Metropolitan University, Tokyo 192-0397}
\affiliation{Virginia Polytechnic Institute and State University, Blacksburg, Virginia 24061}
\affiliation{Wayne State University, Detroit, Michigan 48202}
\affiliation{Yamagata University, Yamagata 990-8560}
\affiliation{Yonsei University, Seoul 120-749}
  \author{M.~Sumihama}\affiliation{Gifu University, Gifu 501-1193}\affiliation{Research Center for Nuclear Physics, Osaka University, Osaka 567-0047} 
  \author{I.~Adachi}\affiliation{High Energy Accelerator Research Organization (KEK), Tsukuba 305-0801}\affiliation{SOKENDAI (The Graduate University for Advanced Studies), Hayama 240-0193} 
  \author{J.~K.~Ahn}\affiliation{Korea University, Seoul 136-713} 
  \author{H.~Aihara}\affiliation{Department of Physics, University of Tokyo, Tokyo 113-0033} 
  \author{S.~Al~Said}\affiliation{Department of Physics, Faculty of Science, University of Tabuk, Tabuk 71451}\affiliation{Department of Physics, Faculty of Science, King Abdulaziz University, Jeddah 21589} 
  \author{D.~M.~Asner}\affiliation{Brookhaven National Laboratory, Upton, New York 11973} 
  \author{H.~Atmacan}\affiliation{University of South Carolina, Columbia, South Carolina 29208} 
  \author{T.~Aushev}\affiliation{Moscow Institute of Physics and Technology, Moscow Region 141700} 
  \author{R.~Ayad}\affiliation{Department of Physics, Faculty of Science, University of Tabuk, Tabuk 71451} 
  \author{V.~Babu}\affiliation{Tata Institute of Fundamental Research, Mumbai 400005} 
% ...
\collaboration{The Belle Collaboration}


\begin{abstract}
We report the first observation of the doubly-strange baryon $\Xi(1620)^0$ in its decay to $\Xi^-\pi^+$ 
via $\Xi_c^+ \rightarrow \Xi^- \pi^+ \pi^+$ decays based on a  $980\,{\rm fb}^{-1}$ data sample 
collected with the Belle detector at the KEKB asymmetric-energy $e^+ e^-$ collider. 
The mass and width are measured to be 1610.4 $\pm$ 6.0 (stat) $^{+5.9}_{-3.5}$ (syst)~MeV$/c^2$ and
59.9 $\pm$ 4.8 (stat) $^{+2.8}_{-3.0}$ (syst)~MeV, respectively.
We obtain 4.0$\sigma$ evidence of the $\Xi(1690)^0$ with the same data sample. 
These results shed light on the structure of hyperon resonances with strangeness $S=-2$.
\end{abstract}


\maketitle

\tighten

{\renewcommand{\thefootnote}{\fnsymbol{footnote}}}
\setcounter{footnote}{0}

The constituent quark model has been very successful in describing the ground state of the
flavor SU(3) octet and decuplet baryons~\cite{PDG,CI,CIK}. 
However, some observed excited states do not agree well with the theoretical prediction.
It is thus important to study such unusual states, both to probe the limitation of the quark models and to spot
unrevealed aspects of the quantum-chromodynamics(QCD) description of the structure of hadron resonances. 
Intriguingly, the $\Xi$ resonances with strangeness $S=-2$ may provide important
information on the latter aspect.

The quantum numbers of several nucleons and $S=-1$ hyperon resonances have been measured.
Recently, there has been significant progress in the experimental study of charmed baryons 
by the Belle, BaBar, and LHCb collaborations.
In contrast, only a small number of $\Xi$ states have been measured~\cite{PDG}.
Neither the first radial excitation with the spin-parity of
$J^P = \frac{1}{2}^+$ nor a first orbital excitation with $J^P = \frac{1}{2}^-$ has been identified. 
Determination of the mass of the first excited state is a vital test of our understanding of the structure of $\Xi$ resonances. 
One candidate for the first excited state is the $\Xi(1690)$, which has a three-star rating on a four-star scale~\cite{PDG}. 
Another candidate is the $\Xi(1620)$, with a one-star rating~\cite{PDG}. 
If the $\frac{1}{2}^-$ state is found, it will be the doubly-strange analogue to the $\Lambda(1405)$ state, 
which has been postulated as a candidate meson-baryon molecular state or a pentaquark~\cite{l1405}.

Experimental evidence for the $\Xi(1620) \rightarrow \Xi\pi$ decay was reported in 
$K^-p$ interactions in the 1970's~\cite{exp1,exp2,exp3}.
The mass and width measurements are consistent but have large statistical uncertainties. 
The most recent experiment, in 1981, has not seen this resonance~\cite{exp4}. 
There is a lingering theoretical controversy about the interpretation of the $\Xi(1620)$ and 
$\Xi(1690)$ states~\cite{gralkin,Miyahara,QM,Oh,lutz,ramos,loring,pervin}, extending from 
their assignment in the quark model to their existence. 
This would be addressed with new high-quality experimental results
for the first excited state with $S=-2$. 
The hadronic decays of charmed baryons governed by the $c \rightarrow s$ quark transition
are a good laboratory to probe these strange baryons. 

In this Letter, we study the decay $\Xi_c^+ \rightarrow \Xi^{*0}\pi^+, \Xi^{*0} \rightarrow \Xi^-\pi^+$  
based on a data sample collected with the Belle detector at the KEKB 
asymmetric-energy $e^+e^-$ (3.5 on 8~GeV) collider~\cite{KEKB}. 
The charge conjugate mode is included throughout this Letter. 
The sample corresponds to an integrated luminosity of 980~fb$^{-1}$. 
The major part of the data was taken at the $\Upsilon(4S)$ resonance; in addition, smaller integrated
luminosity samples were collected off resonance and at the $\Upsilon(1S)$, $\Upsilon(2S)$, $\Upsilon(3S)$, and
$\Upsilon(5S)$.  
We use a Monte Carlo simulation (MC) sample to characterize the mass resolution, detector acceptance, and 
invariant mass distribution in the available phase space. 
The MC samples are generated with EVTGEN~\cite{EVTGEN}, 
and the detector response is simulated with GEANT3~\cite{geant3}.  

The Belle detector is a large-solid-angle magnetic 
spectrometer that consists of a silicon vertex detector (SVD),
a 50-layer central drift chamber (CDC), an array of
aerogel threshold Cherenkov counters (ACC),  
a barrel-like arrangement of time-of-flight
scintillation counters (TOF), and an electromagnetic calorimeter
comprised of CsI(Tl) crystals (ECL); all these components are located inside 
a superconducting solenoid coil that provides a 1.5~T
magnetic field.  
The detector is described in detail elsewhere~\cite{Belle}.
Two inner detector configurations were used. A 2.0 cm radius beampipe
and a 3-layer SVD was used for the first sample
of 156~fb$^{-1}$, while a 1.5 cm radius beampipe, a 4-layer
SVD and a small-cell inner CDC were used to record  
the remaining  824~fb$^{-1}$\cite{svd2}.  

We reconstruct the $\Xi_c^+$ via the $\Xi_c^+ \rightarrow \Xi^- \pi^+ \pi^+,~\Xi^- \rightarrow \Lambda \pi^-,~\Lambda \rightarrow p\pi^-$ decay channel. 
Final-state charged particles, $p$ and $\pi^{\pm}$, are identified using the information 
from the tracking (SVD, CDC) and charged-hadron identification (CDC, ACC, TOF) systems 
combined into likelihood ratios $\mathcal{L}(i:j) = \mathcal{L}_i/(\mathcal{L}_i + \mathcal{L}_j)$, 
where $i,j \in \{p,\,K,\,\pi\}$. 
The $\pi^{\pm}$ particles are selected by requiring the likelihood ratios $\mathcal{L}(\pi:K)>0.6$; this has about $90$\% efficiency. 
The likelihood ratios $\mathcal{L}(p:\pi)>0.6$ and $\mathcal{L}(p:K)>0.6$ are required for proton 
candidates from the $\Lambda$. 
The $\Lambda$ particles are reconstructed from $p\pi^-$ pairs with about $98$\% efficiency. 
The three-momentum of the $\Lambda$ is combined with that of a $\pi^-$ track to reconstruct the helix trajectory of the $\Xi^-$ candidate;
this helix is extrapolated back toward the IP.
A vertex fit is applied to the $\Xi^- \rightarrow \Lambda \pi^-$ decay and the $\chi^2$ 
is required to be less than 50. 
We retain $\Xi^-$ candidates whose mass is within $\pm 3.0\,{\rm MeV}/c^2$($\pm 3\sigma$) of the nominal $\Xi^-$ mass. 
Then, we combine the $\Xi^-$ with two $\pi^+$ candidates, where the pion 
with the lower (higher) momentum is labeled $\pi^+_L$ ($\pi^+_H$). 
The closest distance between the $\pi^+$ track and the nominal $e^+e^-$ interaction point must satisfy  
$|dz| < 1.3~$cm along the beam direction, and 
$|dr| < 0.16~(0.13)~$cm in the transverse plane for $\pi^+_L$~($\pi^+_H$) 
for both $\pi^+_L$ and $\pi^+_H$.
A vertex fit is applied to the $\Xi_c^+ \rightarrow \Xi^-\pi^+\pi^+$ decay. 
The $\chi^2$ is required to be less than 50. 
To purify the $\Xi_c^+$ samples, the scaled momentum $x_p = p_{CM} / \sqrt {\frac{1}{4}s -  m(\Xi_c^+)^2}$  
is required to exceed 0.5, where $p_{CM}$ is the momentum of $\Xi_c^+$ 
in the $e^+e^-$ center-of-mass system, 
$s$ is the squared total center-of-mass energy, and  $m(\Xi_c^+)$ is the $\Xi_c^+$ nominal mass.  
We retain $\Xi_c^+$ candidates that satisfy $|M(\Xi^-\pi^+\pi^+)-m(\Xi_c^+)|< 12.7$~MeV/$c^2$. 
The region 30.0~MeV/$c^2<|M(\Xi^-\pi^+\pi^+)-m(\Xi_c^+)|< $55.4~MeV/$c^2$  
defines the sideband for estimation of the combinatorial background.  

\begin{figure}[htb]
\includegraphics[width=0.5\textwidth]{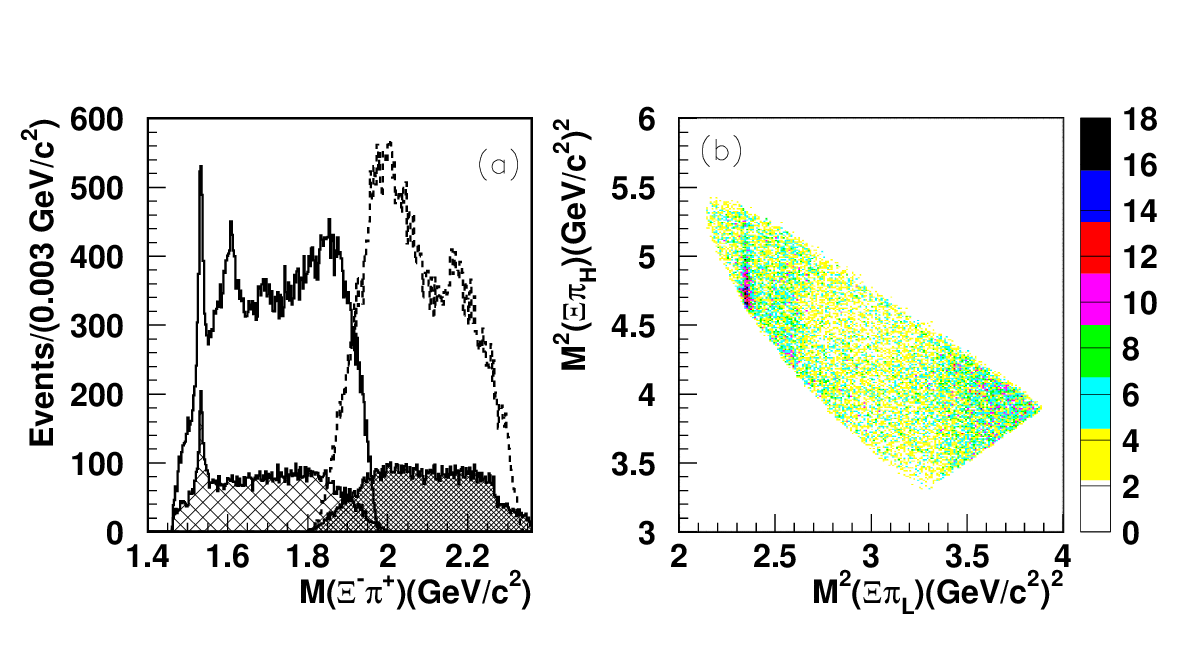}
\caption{(a) The $\Xi^-\pi^+_L$(solid) and $\Xi^-\pi^+_H$(dashed) invariant mass distributions in the $\Xi_c^+$ signal region, 
  as well as the corresponding distributions (hatched) in $\Xi_c^+$ sideband region. (b) The Dalitz distribution for $\Xi_c^+ \rightarrow \Xi^-\pi^+_H\pi^+_L$. (color online)
}
\label{mxipi}
\end{figure}

The $M(\Xi^-\pi^+_L)$ and $M(\Xi^-\pi^+_H)$ distributions of the final sample are shown 
in Fig.~\ref{mxipi}(a). 
Peaks corresponding to $\Xi(1530)^0,~\Xi(1620)^0,$ and $\Xi(1690)^0$ are observed in the 
$M(\Xi^-\pi^+_L)$ distribution.
A reflection due to $\Xi(1530)^0$ decays is seen around 2.2~GeV/$c^2$ in $M(\Xi^-\pi^+_H)$. 
The hatched histograms are the distributions of the $\Xi_c^+$ sideband events, where  
only the $\Xi(1530)^0$ is observed. 
The Dalitz plot of $M^2(\Xi^-\pi^+_L)$ vs. $M^2(\Xi^-\pi^+_H)$ is shown in Fig.~\ref{mxipi}(b). 
The cluster of events due to the $\Xi(1530)^0$ is seen.  
The region 4.3 $-$ 5.3~(GeV/$c^2)^2$ in $M^2(\Xi^-\pi^+_H)$ contains the $\Xi(1620)^0$ and $\Xi(1690)^0$ signals.  
There are currently no known particles with a mass in the range of 2.1 $-$ 2.3~GeV/$c^2$ that would decay into $\Xi\pi$. 
Such massive particles would decay predominantly into a three-particles final state such as $\Xi\pi\pi$. 
The peaks around 1.60 and 1.69~GeV/$c^2$ in $M(\Xi^-\pi^+_L)$ are interpreted as the $\Xi(1620)^0$ and $\Xi(1690)^0$ 
resonances. 
We see an unknown structure in the range 1.8 $-$ 2.1~GeV/$c^2$ in $M(\Xi^-\pi^+)$. 
These events are expected to be due to resonances such as $\Xi(1820)^0,~\Xi(1950)^0$, and $\Xi(2030)^0$.    

The correction of the event-reconstruction efficiency is applied to the mass spectrum. 
To calculate this efficiency,  we generate MC events for the non-resonant three-body decay 
$\Xi_c^+ \rightarrow \Xi^-\pi^+\pi^+$ with a uniform distribution in phase space. 
The efficiency is the number of events surviving the selections   
divided by the total number of generated events, and is measured as a function of
$M(\Xi^-\pi^+_L)$;
the resulting efficiency is from 0.082 to 0.097 and shows a nearly flat distribution in $M(\Xi^-\pi^+_L)$. 
The mass distribution is divided by this efficiency and is normalized by the total number of events.  

\begin{figure}[htbp]
\begin{center}
\begin{tabular}{cc}

\begin{minipage}{.47\hsize}
\begin{center}
\includegraphics[scale=0.2]{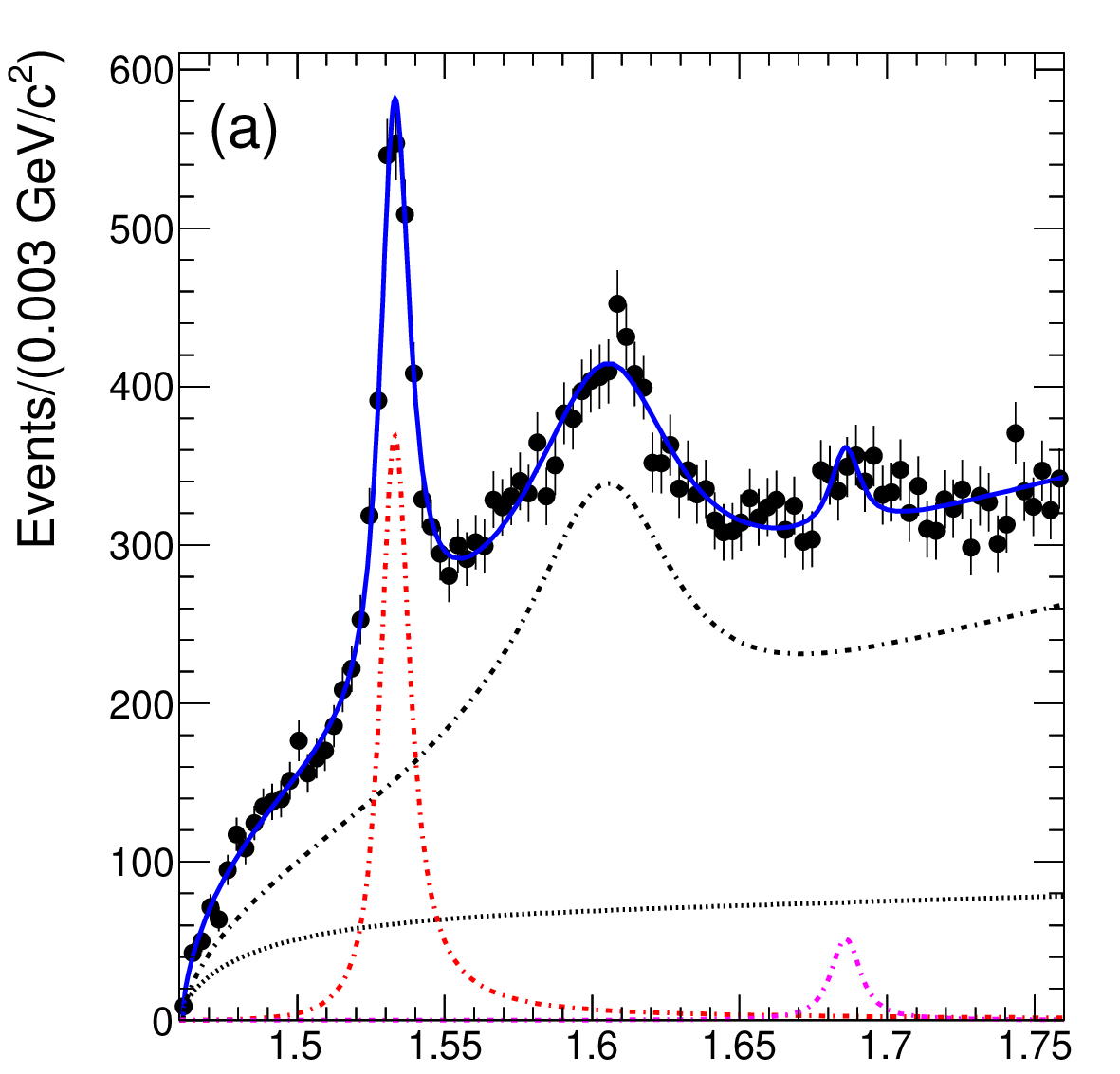}
\includegraphics[scale=0.215]{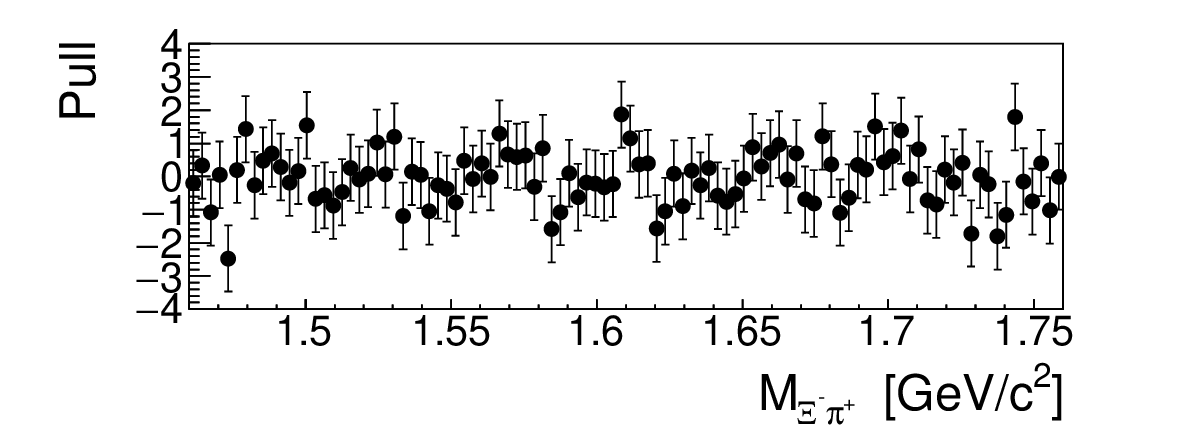}
\end{center}
\end{minipage}

\begin{minipage}{.47\hsize}
\begin{center}
\includegraphics[scale=0.2]{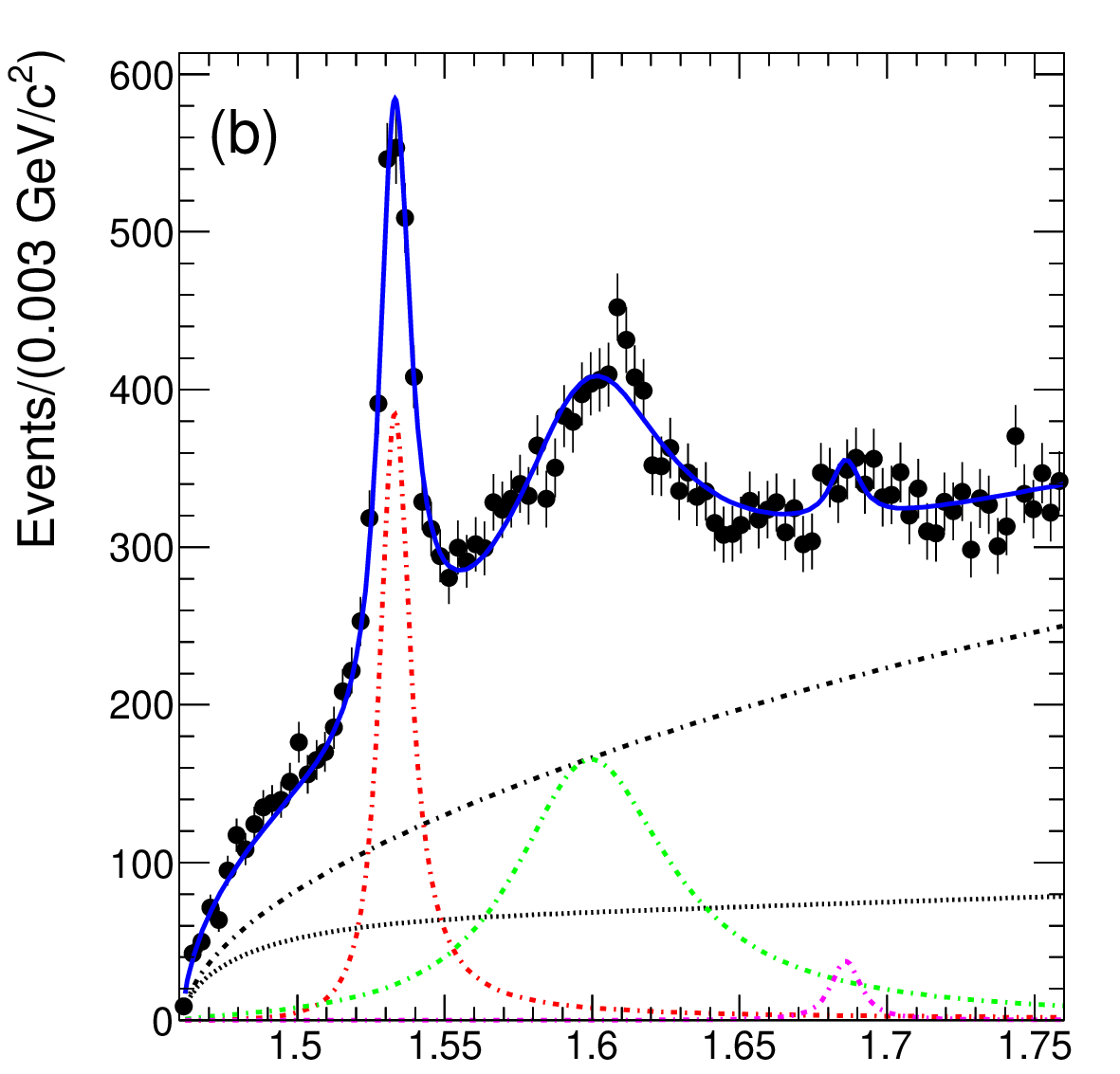}
\includegraphics[scale=0.215]{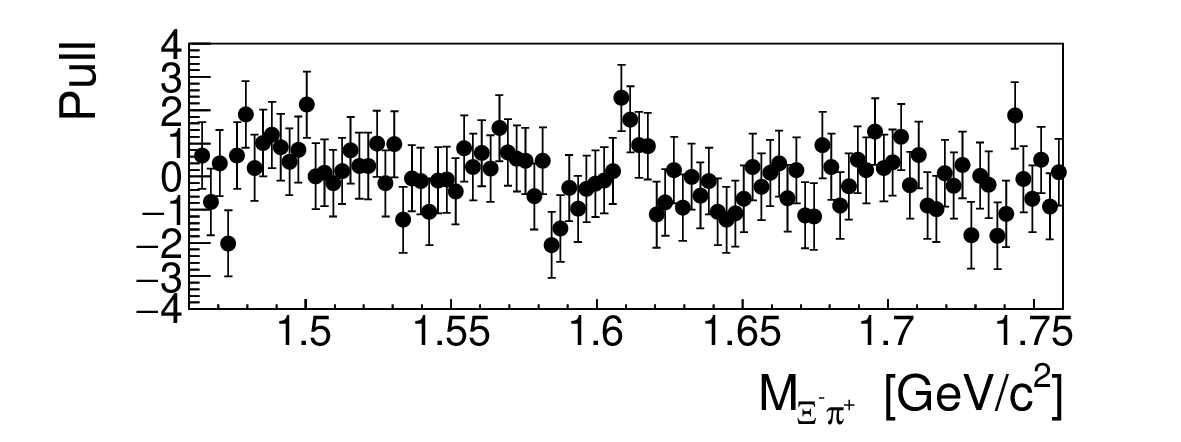}
\end{center}
\end{minipage}

\end{tabular}
\caption{(a) The $\Xi^-\pi^+_L$ invariant mass spectrum (points with error bars),
  together with the fit result (solid blue curve) including the following components:
  $\Xi(1530)^0$ signal (dashed red curve),
  $\Xi(1690)^0$ signal (dot-dashed pink curve),
  $\Xi(1620)^0$ signal and 
  non-resonant contribution (dot-dashed black curve),
  the combinatorial backgrounds (dotted black curve).  
The bottom plots show the normalized residuals (pulls) of the fits.  
(b) The fit without the interference between $\Xi(1620)^0$ and the S-wave non-resonant process. 
The dot-dashed black curve represents the S-wave non-resonant process and
the dot-dashed green curve represents the $\Xi(1620)^0$. (color online)}
\label{fit}
\end{center}
\end{figure}

We perform a binned maximum-likelihood fit to the efficiency-corrected $M(\Xi^-\pi^+_L)$
distribution. 
The fit is applied for the data samples in the signal region and the sideband region simultaneously. 
The fitting range is restricted to (1.46, 1.76)~GeV/$c^2$ to avoid inclusion of the unknown structure between 1.8 and 2.1~GeV/$c^2$.  
The fitting function for the mass spectrum in the signal region includes resonances due to the $\Xi(1530)^0, ~\Xi(1620)^0,$
and $\Xi(1690)^0$, a non-resonant contribution, and the combinatorial background.
The fitting function for the mass spectrum in the sideband region includes the $\Xi(1530)^0$ signal 
and the combinatorial background. 
The shape of the fitting function for the combinatorial backgrounds is common for the mass spectra in the signal region
and the sideband region, and is made by a function with a threshold:
$u^a$exp$(ub)+cu$, where $u=1-[(2-M)/(2-d)]^2$ and $M$ = $M(\Xi^-\pi^+_L)$; $a,b,c,$ and $d$ are free parameters. 
We assume an S-wave non-resonant contribution, and generate the distribution from the MC simulation
of $\Xi_c^+ \rightarrow \Xi^-\pi^+\pi^+$ decays with a uniform distribution in phase space. 
The $\Xi(1620)^0$ signal is modeled with the S-wave relativistic Breit-Wigner function. 
The interference between $\Xi(1620)^0$ and the S-wave non-resonant process is taken into account, and 
these are coherently added. 
The $\Xi(1530)^0$ and $\Xi(1690)^0$ signals are modeled with P- and S-wave 
relativistic Breit-Wigner functions convolved with a fixed Gaussian resolution function of
 width 1.38~MeV/$c^2$ and 2.04~MeV/$c^2$, respectively, as determined from the MC simulation. 
The width and mass of $\Xi(1530)^0$ and $\Xi(1620)^0$ particles are floated in the fit.
The mass and width of the $\Xi(1690)^0$ are fixed in the fit to the values (1686~MeV/$c^2$ and 10~MeV, respectively)
measured by the WA89 Collaboration~\cite{WA89-Xipi}.
Figure~\ref{fit}(a) shows the $\Xi^-\pi^+_L$ mass spectrum with the fitting result. 
The $\chi^2$/ndf (where ndf is the number of degrees of freedom) is 66/86. 
For the $\Xi(1690)^{0}$ resonance, the fit is repeated by fixing the yield to zero; the resulting difference
in log-likelihood with respect the nominal fit and the change of the number of degrees of freedom are used to obtain
the signal significance. 
The statistical significance of the $\Xi(1690)^0$ is 4.5$\sigma$. 
To check the stability of the significance of the $\Xi(1690)^0$, various fit conditions are tried. 
When the P-wave-only relativistic Breit-Wigner with fixed mass and width is used as the fitting function, the significance is 4.0$\sigma$.
When the S-wave-only relativistic Breit-Wigner with the floated mass and width is used, the significance is 4.6$\sigma$.
We take the minimum value of 4.0$\sigma$ as the significance including the systematic uncertainty. 
The measured mass and width of $\Xi(1530)^0$ are 1533.4 $\pm$ 0.35~MeV/$c^2$ and 11.2 $\pm$ 1.5~MeV, respectively.
The measured mass and width of $\Xi(1620)^0$ are 1610.4 $\pm$ 6.0~MeV/$c^2$ and 60.0 $\pm$ 4.8~MeV, respectively.
The mass resolution ($\sigma$) at ~1600~MeV/$c^2$ is 1.6~MeV/$c^2$ as determined from the MC simulation. The width of the $\Xi(1620)^0$ is
59.9~MeV after incorporating this mass resolution. 

We itemize the systematic uncertainties on the mass and width
of the $\Xi(1620)^0$ resonance in Table~\ref{tab:sys}.  
The mass scale and width is checked by comparing the reconstructed mass of the $\Xi(1530)^0$ in the $\Xi^-\pi^+$ 
channel with the nominal mass. 
The differences of the mass and width are $-1.5$~MeV/$c^2$ and $-2.7$~MeV, respectively. 
We then generate and simulate $\Xi_c^+ \rightarrow \Xi^* \pi^+$, $\Xi^* \rightarrow \Xi^- \pi^+$ events 
and analyze these events by the same program as for the real data; 
the mass scale is checked by comparing the reconstructed mass of $\Xi^*$ with the generated mass. 
Here, the difference of the mass is $-0.2$~MeV/$c^2$ and the difference of the width is less than the statistical error.
The systematic uncertainty due to the mass shape of the $\Xi(1620)^0$ is obtained by applying
the fit with the P-wave relativistic Breit-Wigner function instead of the S-wave function. 
The systematic error due to the mass shape of the $\Xi(1690)^0$ is obtained by applying
the fit with the P-wave relativistic Breit-Wigner function instead of the S-wave function, with floated mass and width. 
The nominal bin width of the mass spectrum is 3.0~MeV/$c^2$. 
We determine its systematic uncertainty by changing the bin size from 2.5 to 3.5~MeV/$c^2$ and refitting. 


All of the above sources are uncorrelated, so the total systematic uncertainty is calculated by summing them in quadrature. 
\begin{table}[htbp]
\vspace*{0cm}
\begin{center}
\caption{Systematic uncertainties for the mass and the width of $\Xi(1620)^0$. }
\label{tab:sys}
\begin{tabular}{ccc} \hline \hline
 Source     & \shortstack{Mass\\~(MeV/$c^2$)} &\shortstack{Width\\~(MeV)}  \\ \hline
Mass scale                & $-1.5$        & $-2.7$   \\
Mass shape of $\Xi(1620)$ & $+4.5$        & $+1.8$   \\ 
Mass shape of $\Xi(1690)$ & $+2.3$        & $+1.7$   \\ 
Bin size                  & $\pm3.1$      & $\pm1.3$ \\  \hline
Total                     & $^{+5.9}_{-3.5}$ & $^{+2.8}_{-3.0}$       \\ \hline   \hline  
\end{tabular}
\end{center}
\end{table}

We refit the data using a function that excludes the interference between $\Xi(1620)^0$ and
the S-wave non-resonant process.    
Figure~\ref{fit}(b) shows the $\Xi^-\pi^+_L$ mass spectrum with this hypothesis. 
The $\chi^2$/ndf is 80/87, which is worse than for the nominal fit. 
Here, the measured mass and width of the $\Xi(1620)^0$ 
are 1601.2 $\pm$ 1.5~MeV/$c^2$ and 63.6 $\pm$ 8.7~MeV, respectively.  

For the first time, the $\Xi(1620)^0$ particle is observed in its decay
to $\Xi^-\pi^+$ via $\Xi_c^+ \rightarrow \Xi^- \pi^+ \pi^+$ decays. 
The number of $\Xi(1620)^0$ events is two orders of magnitude larger than in previous experiments.
The measured mass and width of the $\Xi(1620)^0$ are consistent with
the results of previous measurements within the large uncertainties
of the latter and are much more precise.
The width of the $\Xi(1620)^0$ is somewhat larger than that of the other
$\Xi^*$ particles~\cite{PDG}.

The constituent quark models have predicted the first excited states of $\Xi$ around 1800 MeV/$c^2$~\cite{CI}; 
therefore, it is difficult to explain the structure of the $\Xi(1620)^0$ and $\Xi(1690)^0$
in this context.
Instead, it implies that these states are candidates of a new class of exotic hadrons.
We observe in the low-mass region two states with
a mass difference of about 80 MeV/$c^2$: the $\Xi(1620)^0$ is
strongly coupled to $\Xi\pi$ and the $\Xi(1690)^0$ to $\Sigma K$.
The situation is similar to the two poles of the $\Lambda(1405)$~\cite{l1405} and suggests 
the possibility of two poles in the $S=-2$ sector.
Studying these states may explain the riddle about the $\Lambda(1405)$; consequently,  
the interplay between the $S=-1$ and $S=-2$ states can help resolve this longstanding
problem of hadron physics.

The $\Xi(1620)^0$ and $\Xi(1690)^0$ particles are found in the decay of $\Xi_c^+$ while   
their signals are not seen in the sideband events of Fig.\ref{mxipi}(a). 
These results offer a clue for understanding the quark structure of these exotic states. 
The result indicates that 
the hadronic decays of charmed baryons via charm-to-strange quark transitions are potentially a promising  system for
further studies of strange baryons~\cite{Miyahara}. 




We thank the KEKB group for the excellent operation of the
accelerator; the KEK cryogenics group for the efficient
operation of the solenoid; and the KEK computer group, and the Pacific Northwest National
Laboratory (PNNL) Environmental Molecular Sciences Laboratory (EMSL)
computing group for strong computing support; and the National
Institute of Informatics, and Science Information NETwork 5 (SINET5) for
valuable network support.  We acknowledge support from
the Ministry of Education, Culture, Sports, Science, and
Technology (MEXT) of Japan, the Japan Society for the 
Promotion of Science (JSPS), and the Tau-Lepton Physics 
Research Center of Nagoya University; 
the Australian Research Council including grants
DP180102629, 
DP170102389, 
DP170102204, 
DP150103061, 
FT130100303; 
Austrian Science Fund under Grant No.~P 26794-N20;
the National Natural Science Foundation of China under Contracts
No.~11435013,  
No.~11475187,  
No.~11521505,  
No.~11575017,  
No.~11675166,  
No.~11705209;  
Key Research Program of Frontier Sciences, Chinese Academy of Sciences (CAS), Grant No.~QYZDJ-SSW-SLH011; 
the  CAS Center for Excellence in Particle Physics (CCEPP); 
the Shanghai Pujiang Program under Grant No.~18PJ1401000;  
the Ministry of Education, Youth and Sports of the Czech
Republic under Contract No.~LTT17020;
the Carl Zeiss Foundation, the Deutsche Forschungsgemeinschaft, the
Excellence Cluster Universe, and the VolkswagenStiftung;
the Department of Science and Technology of India; 
the Istituto Nazionale di Fisica Nucleare of Italy; 
National Research Foundation (NRF) of Korea Grants
No.~2015H1A2A1033649, No.~2016R1D1A1B01010135, No.~2016K1A3A7A09005
603, No.~2016R1D1A1B02012900, No.~2018R1A2B3003 643,
No.~2018R1A6A1A06024970, No.~2018R1D1 A1B07047294; Radiation Science Research Institute, Foreign Large-size Research Facility Application Supporting project, the Global Science Experimental Data Hub Center of the Korea Institute of Science and Technology Information and KREONET/GLORIAD;
the Polish Ministry of Science and Higher Education and 
the National Science Center;
the Grant of the Russian Federation Government, Agreement No.~14.W03.31.0026; 
the Slovenian Research Agency;
Ikerbasque, Basque Foundation for Science, Basque Government (No.~IT956-16) and
Ministry of Economy and Competitiveness (MINECO) (Juan de la Cierva), Spain;
the Swiss National Science Foundation; 
the Ministry of Education and the Ministry of Science and Technology of Taiwan;
and the United States Department of Energy and the National Science Foundation.

\newcommand{\divergence}{\mathrm{div}\,} 
\newcommand{\grad}{\mathrm{grad}\,} 
\newcommand{\rot}{\mathrm{rot}\,}  
\newcommand{\etal}{{\em et al.}} 
\def\Journal#1#2#3#4{{#1} {\bf #2} #4 (#3)}
\def\NCA{\em Nuovo Cimento}
\def\NIM{Nucl. Instrum. Methods}
\def\NIMA{Nucl. Instrum. Meth. Phys. Res. A}
\def\NPA{Nucl. Phys. A}
\def\NPB{Nucl. Phys. B}
\def\PREP{Phys. Pep.}
\def\PLB{Phys. Lett.  B}
\def\PRL{Phys. Rev. Lett.}
\def\PRC{Phys. Rev. C}
\def\PRD{Phys. Rev. D}
\def\PRG{Prog. Part. Nucl. Phys.}
\def\PPNP{Progr. Part. Nucl. Phys.}
\def\RMP{Rev. Mod. Phys.}
\def\ZPC{Z. Phys. C}
\def\PTEP{Prog. Theor. Exp. Phys.}
\def\EPJ{Eur. Phys. J.}

\end{document}